# Unfolding the Neutron Spectra from a Water-Pumping-Injection Multi-layered Concentric Sphere Neutron Spectrometer Using a Self-Adaptive Differential Evolution Algorithm

Rui Li[1]　Jianbo Yang[2,*]　Xianguo Tuo[2]　Rui Shi[2]　Jie Xu[1]

**Affiliations:**

[1]Chengdu University of Technology, Chengdu 610059, China

[2]Sichuan University of Science and Engineering, Zigong 643000, China

[*]Corresponding author. *E-mail address:* yjb@cdut.edu.cn

**Abstract:** A self-adaptive differential evolution neutron spectrum unfolding algorithm (SDENUA) was established in this paper to unfold the neutron spectra obtained from a Water-pumping-injection Multi-layered concentric sphere Neutron Spectrometer (WMNS). Specifically, the neutron fluence bounds were estimated to accelerate the algorithm convergence, the minimum error between the optimal solution and the input neutron counts with relative uncertainties was limited to $10^{-6}$ to avoid useless calculation. Furthermore, the crossover probability and scaling factor were controlled self-adaptively. FLUKA Monte Carlo was used to simulate the readings of the WMNS under (1) a spectrum of Cf-252 and (2) its spectrum after being moderated, (3) a spectrum used for BNCT, and (4) a reactor spectrum, and the measured neutron counts unfolded by using the SDENUA. The uncertainties of the measured neutron count and the response matrix are considered in the SDENUA, which does not require complex parameter tuning and the priori default spectrum. Results indicate that the solutions of the SDENUA are more in agreement with the IAEA spectra than that of the MAXED and GRAVEL in UMG 3.1, and the errors of the final results calculated by SDENUA are under 12%. The established SDENUA has potential applications for unfolding spectra from the WMNS.

## 1 Introduction

Since the first introduction of Bonner Sphere Spectrometer (BSS) in 1960 [1], it has been widely used in neutron spectrometry measurements, such as the isotopic neutron source [2], the BNCT [3], and radiation protection near the reactor [4], because of its advantages in isotropic response and wide energy range. A newly designed neutron spectrometer, the Water-pumping-injection Multi-layered concentric sphere Neutron Spectrometer (WMNS) uses water as a moderator [5-7], and the principle of neutron spectrometry measurement is similar to that of BSS. The structure of WMNS is shown in Fig. 1. Seven stainless-steel spherical shells are arranged concentrically to build six spherical gaps, five of them are used to hold water (the thickness of the water gap from outside to inside is 2.5, 3.75, 8, 1.25, and 1 cm in order), and the rest one gap is filled with lead. The water will be independently pumped into each gap to form a measurement unit (combination) to moderate the incident neutrons, and an easy-to-replace spherical $^3$He proportional counter (model: LND 2705) is placed in the innermost to detect the thermal neutron. 5 gaps have up to 32 measurement combinations with different thickness water (similar to the ball with different diameters in the BSS), so, 32 measurement combinations that can be used to collect neutron count. The switching between measurement combinations is realized by using the external water-pumping-injection system [8]. The 1 cm lead is utilized, which aims to measure the high-energy neutron. WMNS is a portable and flexible neutron spectrometer. An active or passive detector can be chosen depending on the measurement environment, and just one reading electric system is required, and the "ball-ball interference" problem in the traditional BSS is eliminated by integrating the multi-concentric-spheres. The measurement combinations are coded by 0 and 1, for example, the combination code of Fig. 1 is "00Pb110", where "0" is the gap filled with air, "1" is the gap filled with water, and "Pb" is the lead.

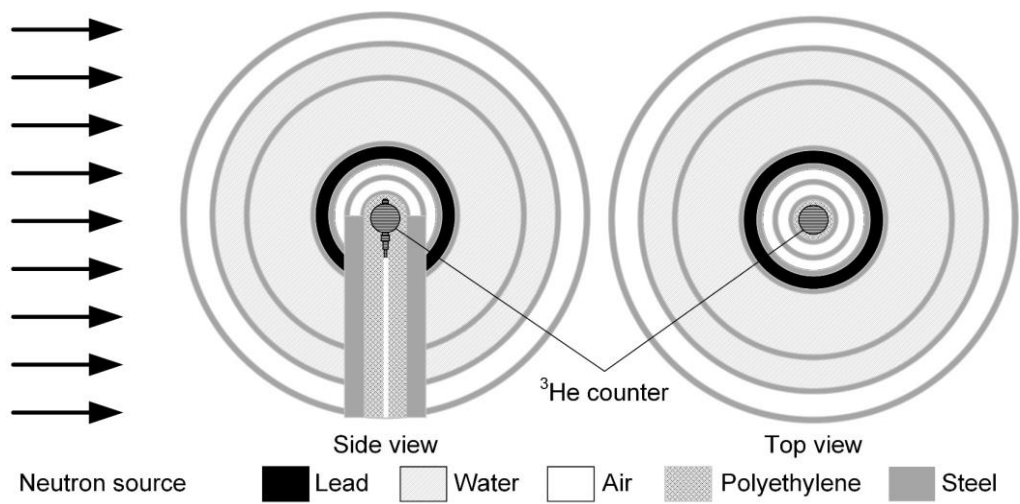

Fig. 1 The schematic diagram of the WMNS

The readings of the ³He proportional counter also called measured counts, are the nuclear reaction event counts of ³He(n, p)³H under different measurement combinations. The target spectrum is unfolded from these measured counts using the neutron unfolding algorithm, and the neutron unfolding process could be presented in a discrete form as [9]

$$C_j^{meas} + \varepsilon_j = \sum_{i=1}^{n} R_{ij}\varphi_i \quad j=1,2,3,...m, \quad (1)$$

where $C_j^{meas}$ is the measured neutron count reading from the $j$th measurement combination, $\varepsilon_j$ is the reading uncertainty of the $j$th measurement combination, $R_{ij}$ is the response of the $j$th measurement combination to the neutron of $i$th energy group, and $\varphi_i$ is the neutron fluence of $i$th energy group. Normally, the number of measurement combinations is far smaller than the number of energy groups.

In WMNS, to minimize the time taken for switching measurement combinations, 18 measurement combinations were selected. 36 energy groups were divided logarithmically at equal intervals with the range from $10^{-9}$ MeV to 20 MeV, which aimed to reduce the underdetermined degree of the unfolding process. So, m = 18, n = 36. The FLUKA Monte Carlo code [10] was used to calculate the response matrix, as shown in Fig. 1, parallel monoenergetic neutron beams, starting from a disk with a diameter of 28 cm, which was the same as the diameter of the outermost stainless-steel shell of WMNS, irradiated the WMNS to obtain responses. The distance between the source and the center of the spectrometer was 60 cm. The RESNUCLEi card was employed to score stopping nuclei on ³He. Stopping nuclei are tritium nuclei and protons because of the ³He(n, p)³H reaction; then, half of all nuclei were collected as the reading of detector because when each tritium nucleus or proton was produced, a neutron would be detected simultaneously. The thermal neutron scattering data S(α, β) was applied to the transport of neutrons below 4 eV in polyethylene and water by using the LOW-

NEUT card and LOW-MAT card [11].

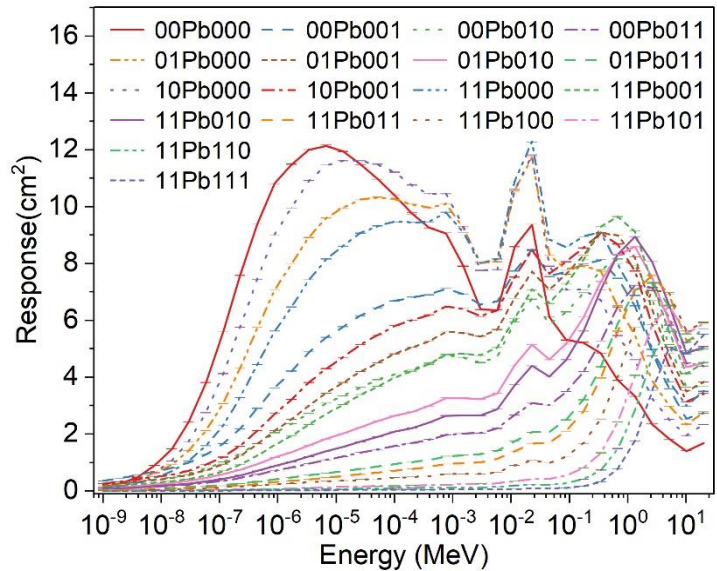

**Fig. 2** The response matrix of the WMNS

At present, various methods, such as the maximum entropy method [12] and iterative method [13] used in the Unfolding with Maxed and Gravel 3.1 (UMG 3.1) [14], can be used to unfold neutron spectrum, and their solutions are compared with that of the present work. An excellent priori default spectrum is required when using UMG 3.1 to unfold neutron energy spectra [12] because for the maximum entropy method, a priori default spectrum is a benchmark for UMG 3.1 deciding when to output the solution, and for the iterative method, the priori default spectrum will be the initial of the iteration. Shahabinejad et al. [15] used a two-step genetic algorithm (TGA) to unfold neutron energy spectra, and the results have shown closer match in all energy regions and particularly in the high energy regions than the common genetic algorithm (GA). Energy groups in the high energy range were unfolded at the first step, which were used to construct the initial value of the second step. After a year, they used a particle swarm optimization algorithm (PSOA) [16] to unfold the neutron spectrum from a pulse height distribution and a response matrix, and the results demonstrated to match well with the TGA. In the PSOA, acceleration constants $c_1$, $c_2$, and inertia weight $w$ were empirically predefined by authors. Hoang et al. [17] applied a different two-step GA to unfold neutron spectra obtained from activation foils. Different from the literature [15], first step, only the energy groups in the region from 20 MeV to 35 MeV were unfolded, and in the second step, the entire energy spectrum was unfolded with keeping the result of the first step constant. K. Chang et al. [6] established a backpropagation artificial neural network neutron spectrum unfolding code, and the training of the neural network was performed under 32 neutron spectra, and its ability was verified by 8 neutron spectra. As mentioned above, the methods in UMG 3.1 rely on a priori default spectrum, parameters tuning in GA and PSOA frameworks are complicated, and the neural network training is also a time-consuming and complex task.

In this paper, we focus on the need to unfold the neutron spectrum from the WMNS. The self-adaptive differential evolution neutron spectrum unfolding algorithm (SDENUA) is proposed, which includes the neutron fluence bounds estimation and parameters self-adaptive control technique. The error between the input neutron counts and the calculated counts is limited to $10^{-6}$, which helps to improve the quality of solutions and reduce the calculation time. The measured neutron counts of 1) spectrum of Cf-252 and 2) its spectrum after being moderated, 3) a spectrum used for BNCT, and 4) a spectrum from a reactor in the IAEA 403 report [18] were simulated by using the FLUKA code, and the measured neutron counts with relative uncertainties were unfolded by using the SDENUA. The SDENUA does not require complex parameter tuning and the priori default spectrum, and the established SDENUA has potential applications for unfolding spectra from the WMNS.

The rest of this paper is arranged as follows. In Section of the Material and methods, firstly, estimation of neutron fluence bounds is performed, secondly, the techniques related to self-adaptive differential evolution algorithm using for neutron energy spectra unfolding are explained in detail along with the formulation of the algorithm, thirdly, the termination criterion of the algorithm is proposed. In Section of the Results and discussion, the unfolded spectra of our work are shown, and compared with the UMG 3.1, the uncertainties of the unfolded spectra are discussed. Finally, the paper is concluded in Section of the Conclusion.

## 2 Material and methods

### 2.1 Estimation of neutron fluence bounds

In the optimization problem, it is necessary to pre-estimate the search space bound of each variable, because the scale of the search space has a significant impact on the running time and convergence of the algorithm [9]. In other words, the neutron fluence of each energy group needs to be bounded before unfolding. As we all know, the actual neutron energy spectrum due to its physical properties, the minimum neutron fluence values of all energy groups are 0, however, the upper fluence needs to be estimated. In Ref. [20], it was assumed that the measured count of a particular measurement unit is fully contributed by a particular energy group, and the minimum fluence value of these estimates is the upper bound of the energy group, as shown in Eq. (3). For the ideal monochromatic pulse neutron energy spectrum, this method can directly give the neutron fluence where the pulse is located. Although this method is based on strict mathematical derivation, since the contribution of the fluence outside the particular energy group to the neutron count is ignored, the result of this method is rough.

The neutron energy spectrum is usually continuous [9], we assume that the fluence changing between adjacent energy groups is relatively smooth. So, the range estimated by the above method is narrowed by

$$\varphi_i^{max} = \frac{\varphi_i^1 + \varphi_i^3}{2}, (2)$$

$$\varphi_i^1 = MIN(C_j^{meas}/R_{ij}), (3)$$

$$\varphi_i^3 = \frac{\varphi_i^1 \cdot \varphi_{\lceil i/3 \rceil}^1}{\sum_i^{i+2} \varphi_i^1}, (4)$$

where $\varphi_i^{max}$ is the fluence upper bound of the $i$th energy group, $\varphi_i^1$ is the fluence upper bound of the $i$th energy group estimated by using the method in Ref. [20], $\varphi_i^3$ is the fluence upper bound of the $i$th energy group with three energy groups, $\lceil i/3 \rceil$ rounds to the nearest integer greater than or equal to $i/3$, that is the first of every three $\boldsymbol{\varphi}^1 = [\varphi_1^1, \varphi_2^1, ...\varphi_i^1...\varphi_n^1]$ ($n$ is the number of the energy group), $C_j^{meas}$ and $R_{ij}$ are same as the Eq. (1). Finally, the fluence upper bound of the $i$th energy group is on the interval $\varphi_i \in (0, \varphi_i^{max})$.

It is worth noting that a smooth neutron spectrum is a prerequisite, if not so, the fluence of the peak will be underestimated. Therefore, it does not recommend estimating the fluence upper bound of the energy spectrum which contains a sharp peak.

**2.2 Self-adaptive differential evolution algorithm**

The Differential Evolution Algorithm (DEA), which was introduced by R. Storn et al. in 1996 [19]. This algorithm captures wide attention and application because of its simple framework and powerful global search capabilities, and the agreement between the individuals and the solutions is evaluated by the fitness value generated by the fitness function. The iterative loop of the algorithm includes evolutionary operations, such as initialization, mutation, crossover, and selection. For neutron unfolding, the algorithm population is composed of several individuals. The individual corresponds to the neutron spectrum composed of several genes. The positions of the genes correspond to the positions of the energy group, and the values of the genes correspond to the neutron fluence, which is to the variables of the unfolding problem to be solved. The SDENUA will be introduced in detail next.

**Initialization** of neutron fluence of each energy group was achieved by randomly selecting from the neutron fluence estimation interval $(0, \varphi_i^{max})$. A key issue in initialization is how many individuals are to be included in the population. Too-small population size can result in premature convergence, however, too many individuals in the population will lead to a long calculation time and the population is too large to get enough mixing. The results of the 20-dimensional (20 variables) problem were examined by R. Gämperle et al. [26], which states that a reasonable choice for the population size is between 3 to 8 times the number of variables, while in Ref. [21], 10 times is recommended. To ensure that the population has enough different vectors to participate in the mutation, and to improve the search and traversal ability of the population in the evolutionary process, the population size in this work was set to 10 times the number of the energy group (*PopSize* = 10×36), and population size remained the same throughout the entire unfolding process. Such a population size will imply a larger number of calculations, so, a time-saving technique will be proposed in section 2.3.

**The mutation** operation is executed as Eq. (5) [22], which guides the direction of evolution of the population. The search step length is controlled by the scaling factor. Therefore, the mutation operation provides two functions of search direction and search step length control evolution.

$$\mathbf{v}_{i,g} = \mathbf{k}_{i,g} + F_i \cdot (\mathbf{x}_{best,g}^{B*PopSize} - \mathbf{k}_{i,g}) + F_i \cdot (\mathbf{x}_{r,g} - \mathbf{x}_{f,g}), \quad (5)$$

where $\mathbf{v}_{i,g}$ is the *i*th temporary individual in the *g*th generation, $\mathbf{k}_{i,g}$ is the *i*th target individual in *g*th generation, $\mathbf{x}_{best,g}^{B*PopSize}$ is the High-fitness individual. $\mathbf{x}_{r,g}$ is randomly selected from the current population $\mathbf{P}$, and $\mathbf{x}_{f,g}$ is randomly selected from ($\mathbf{P}_f \cup \mathbf{P}$), the $\mathbf{P}_f$ is a set used for saving failure individuals (the individuals with lower fitness in the selection step). $F_i$ is the scaling factor of each target individual.

Before the mutation operation, all the individuals in the current population $\mathbf{P}$ are sorted in descending order according to their fitness. High-fitness individuals are randomly selected from the top 100*B% individuals after sorting, and B is a uniform random number on the interval [0.05, 0.60]. The upper bound at 0.6 of B is more suitable for the 36-dimension unfolding problem, which can make more high-fitness individuals to participate in the mutation step and can reduce the risk of premature convergence. The set $\mathbf{P}_f$ is created with the size of 100 from the beginning of the first iteration, the individuals with lower fitness in the selection step would be sorted in the $\mathbf{P}_f$, When the $\mathbf{P}_f$ is full, the failure individuals in the $\mathbf{P}_f$ are updated following the "first in, first out" rule. The $\mathbf{P}_f$ holds failure individuals from 100 generations to provide more different genes to mutation.

Each scaling factor is generated independently according to a normal distribution, that is, $F_i \sim N(u_F, 0.1)$, which can give better results comparing with the Cauchy sampling method, and $u_F$ is updated [22] at the end of each generation by

$$u_F = (1-c) \cdot u_F + c \cdot \text{mean}_L(S_F), \quad (6)$$

where $\text{mean}_L(\cdot)$ is the Lehmer mean

$$\text{mean}_L(S_F) = \frac{\sum_{F \in S_F} F^2}{\sum_{F \in S_F} F}, \quad (7)$$

where the $S_F$ is a set for all failure mutation factors in each generation, and the $u_F$ saved in the $S_F$ are used to provide the prior historical information for the next $u_F$ updating. $S_F$ would be blanked at the beginning of each generation. $c$ is a uniform random number on the interval [0.05, 0.20], which controls the life span of $u_F$ with generations from 5 to 20 [22], and $u_F = 0.5$ at the initialization。

**The crossover operation** is based on the temporary individuals generated by the mutation operation, and makes them crossover with the target individuals as

$$u_{i,g}^j = \begin{cases} v_{i,g}^j, & r^j \leq CR_i \\ k_{i,g}^j, & \text{otherwise} \end{cases}, \quad (8)$$

where $u_{i,g}^j$ is the $j$th gene of the $i$th candidate individual, $v_{i,g}^j$ is the $j$th gene of $\mathbf{v}_{i,g}$, $k_{i,g}^j$ is the $j$th gene of $\mathbf{k}_{i,g}$, $r^j$ is a uniform random number on the interval [0, 1], and $CR_i$ is the crossover probability of each candidate individual.

The basic unit of crossover is the gene, the genes in the temporary individual and the target individual are extracted orderly to construct a candidate individual according to the crossover probability. The candidate individuals indirectly transmit the impact from the mutation operation on the target individual, and the impact is controlled by the crossover probability. The self-adaptive adjustment of crossover probability based on the historical evolution information is more effective than the traditional constant control method.

Each $CR_i$ is generated as Eq. (9) and updated as Eq. (10) [22]

$$CR_i = \begin{cases} CR_i \sim N(u_{CR}, 0.1), \text{otherwise} \\ CR_i = 0.5, \quad CR_i < 0.5 \\ CR_i = 0.95, \quad CR_i > 0.95 \end{cases}, (9)$$

$$u_{CR} = (1-c) \cdot u_{CR} + c \cdot S_{CR}^{Mean}, (10)$$

where $S_{CR}^{Mean}$ is the average of all elements in the set $S_{CR}$, each $CR_i$ would be sent into the set $S_{CR}$ if the fitness of the target individual higher than the candidate individual in the selection operation. $S_{CR}$ would be blanked at the beginning of each iteration. $c$ is a uniform random number on the interval [0.05, 0.20], which controls the life span of $u_F$ with generations from 5 to 20 [22], and $u_F = 0.5$ at the initialization [19, 22]. To ensure that the genes both of the temporary and the target individuals could be passed partly on to the candidate individuals, $CR_i$ is truncated to [0.5, 0.95] [23, 24]. The candidate individual is generated by the crossover operation will be sent to the selection operation.

**The selection operation** will decide whether to refuse or allow the candidate individual to enter the population based on

$$\mathbf{k}_{i+1,g} = \begin{cases} \mathbf{u}_{i,g}, f(\mathbf{u}_{i,g}) > f(\mathbf{k}_{i,g}) \\ \mathbf{k}_{i,g}, \text{otherwise} \end{cases}, (11)$$

where $f(\cdot)$ is the fitness function. If the candidate individuals with lower fitness, which will be rejected from the population to maintain the average fitness of the population at a higher level. In that case, the efforts of mutation operation and crossover operation will be futile if we do not take advantage of the candidate individuals.

**The fitness function** [17] used to evaluate how close between the input neutron counts and the individuals (solutions), that is,

$$f = [\sum_{j=1}^{m} \frac{(C_j^u - C_j^{cal})^2}{(C_j^u)^2}]^{-1}, (12)$$

where $C_j^u$ is the input neutron count of $j$th measurement combination generated with a relative uncertainty, $C_j^{cal} = \sum_{i}^{n} R_{ij}^u \varphi_i^{cal}$ is the calculated neutron count of $j$th measurement combination, $R_{ij}^u$ is a response generated with relative a uncertainty, $\varphi_i^{cal}$ is the neutron fluence of $i$th energy group of the calculated spectrum (solution). The solution is closer to the $C_j^u$, the fitness is higher. In the spectrum unfolding process, the fitness function is the only criterion for judging the quality of the solution. So, the performance of the fitness functions can have a significant effect on the solution, even the performance of the algorithm. To reduce the dependence of the algorithm on the fitness function, and the fitness value of the optimal solution is restricted to obtain a physically acceptable solution in the next section.

## 2.3 Termination criterion

The neutron energy spectrum unfolding problem is a first kind Fredholm integral problem, it is impossible to obtain a perfect solution based on the integral fitness function as Eq. (12), and a fundamental hypothesis for the unfolding algorithm is that solutions with acceptable spectral quality can be found based on the fitness function [17]. Besides, to improve the probability of convergence and obtain high-fitness solutions, a larger maximum iteration number is usually made to truncate the fitness of the final solution, such as the literature [9, 15], the authors empirically defined a larger maximum iteration number for spectrum unfolding. However, the overestimation of the maximum iteration number would increase calculation time, while the underestimation would output the pseudo optimal solution before converging. So, a better termination criterion is proposed in this work.

A spectrum quality factor (QS) [9] is used to evaluate the quality of the solution as

$$QS = 100 \cdot \sqrt{\frac{\sum_{i=1}^{n}\left(\varphi_i^{ref} - \varphi_i^{cal}\right)^2}{\sum_{i=1}^{n}(\varphi_i^{cal})^2}}, \quad (13)$$

where $\varphi_i^{ref}$ is the neutron fluence of $i$th energy group of the reference spectrum (solution), and $\varphi_i^{cal}$ is the same as Eq. (12). A perfect solution results in QS = 0.

In the unfolding process, the quality of the optimal solution is expected to improve with the fitness of the solutions increasing, which means that the QS of the solutions conforms to the monotonic non-increasing trend. The relationship between QS and fitness of the spectrum "For BNCT" is shown in Fig. 3, with the fitness of the solutions increasing, the QS of the solutions fluctuates. Although the fitness of the final solution is super high, the QS of the final solution is not the lowest in history. In other words, the optimal solution is missed, and the calculation makes a QS rebound after the generation with the lowest QS, which is useless even harmful. However, QS is an evaluation indicator based on a known energy spectrum. In the actual energy spectrum unfolding process, we can only use fitness to evaluate the solutions. To make the QS the final solution as close as possible to the lowest value, after many experiments, the fitness of the optimal solution is limited to $10^6$ as the iteration termination criterion to replace the maximum iteration number as the iteration termination criterion, which helps save the running time and increase the quality of the final optimal solution.

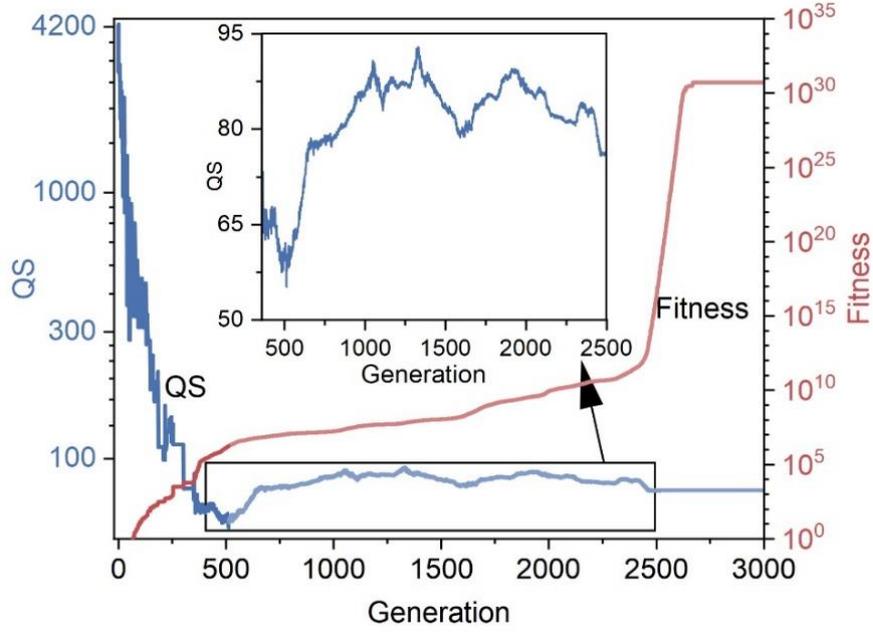

**Fig. 3** The relationship between QS and fitness of the spectrum "For BNCT"

To investigate how uncertainties from the measured neutron counts and from the response matrix affected the final results. The inputs of each run in the SDENUA were generated by

$$C_j^u = C_j^{meas} + rand(-\varepsilon_j,\ \varepsilon_j)C_j^{meas}, (14)$$

$$R_{ij}^u = R_{ij} + rand(-\sigma_{ij},\ \sigma_{ij})R_{ij}\ , (15)$$

where $C_j^u$ and $R_{ij}^u$ are the measured neutron count with uncertainty estimation and the response matrix with uncertainty estimation respectively, $C_j^{meas}$, $R_{ij}$, and $\varepsilon_j$ are the same as Eq. (1), $\sigma_{ij}$ is the uncertainty of each response, and rand(·) is the uniform random sampling function. The maximum errors of measured neutron count simulated from relative reference spectra are 0.37%, 0.75%, 2.23%, and 3.10%. The average and the maximum error of the response matrix are 0.75% and 10.23% respectively. The flow of SDENUA is shown in Fig. 4.

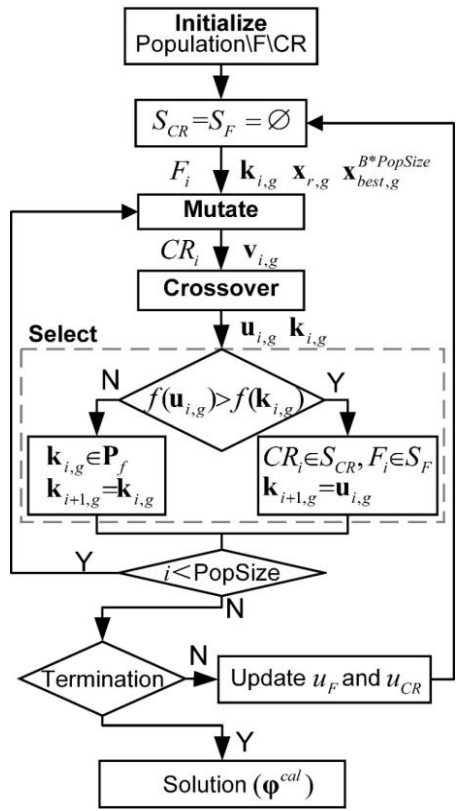

**Fig. 4** The flow of SDENUA

## 3 Results and discussion

As shown in Fig. 5, after multiple independent runs, the QS of the solutions with an average over of 20 independent runs is lower excepted the "Reactor" spectrum. So, 20 times should be a reasonable choice in actual applications.

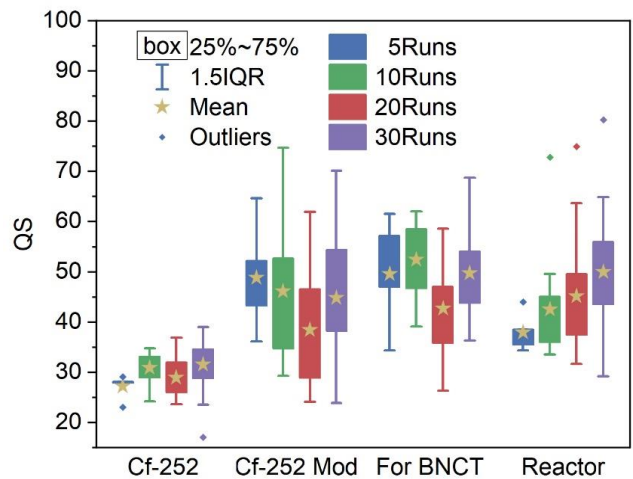

Fig. 5 The QS of the solutions with average over multiple independent runs in SDENUA

As shown in Fig. 6, the termination generations of the final optimal solution of each spectrum are quite different over 20 runs when the fitness of the solutions reached $10^6$, and the average termination generation over 20 runs of "Cf-252" spectrum is 1157th generation, "Cf-252 Mod" spectrum is 544th, "For BNCT" spectrum is 451th, and "Reactor" spectrum is 472th. The termination generations suitable for "Cf-252" and "Cf-252 Mod" are an overestimation as for "For BNCT" and "Reactor", and the opposite is an underestimation. So, there may not be a universal maximum number of iterations. It is proved once again from the other side of the view, that in the neutron spectrum unfolding problem, especially in the face of multiple energy spectrum types, using the maximum number of iterations as the termination condition is not an optimal choice.

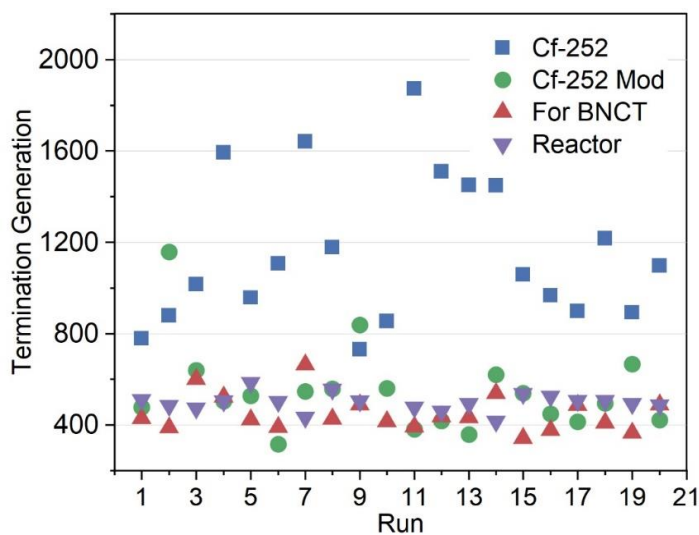

**Fig. 6** Termination generation of the final optimal solution of 20 runs when the fitness of the solutions reaches $10^6$

The termination generations vary so widely between runs and more generations are required in the case of the Cf-252 spectrum. As shown in Fig. 8a, there are many energy groups with 0 fluences. To ensure a small error between the input and calculated counts, the algorithm can only make the fluences close to 0 in the side positive range, because the negative fluences will be rejected. This also means that the solution space that can meet the error constraint is narrowed, and more iterations are required.

Moreover, only the energy groups with non-zero fluences can contribute to neutron counts when the solution is convolved with the response matrix. The fewer non-zero terms in the Cf-252 solution, the more instability of the Cf-252 solution, so, the distribution of solutions will expand.

The oscillations in the unfolded spectra are an inherent feature of the numerical solution to the Fredholm equation [14, 15, 24], and a Gaussian smoothing method was adopted to smooth the final optimal solution to overcome the oscillations. In Gaussian smoothing, the smooth window size is 7, and the sigma is 1.4.

As shown in Fig. 8, the neutron fluence at the peak position of solutions of the SDENUA are underestimated, one of the reasons is the optimal solution of SDENUA smoothed by the Gaussian method artificially increases the error. As shown in Fig. 7, however, the results show that the positive effect of Gaussian smoothing is greater than the error caused by smoothing.

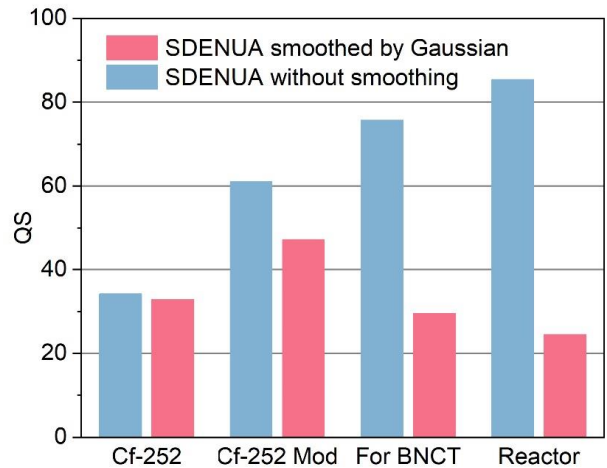

**Fig. 7** The QS of the final solutions from the SDENUA smoothed by using the Gaussian smoothing method or not

As described above in the introduction section, the methods in UMG 3.1 start with a priori default spectrum, to make the comparison more fairly, an excellent priori spectrum was given when using the UMG 3.1 unfolding code. The inputs ($C_j^u$ and $R_{ij}^u$) were also generated as Eq. (14) and Eq. (15), and chi-square was set to $10^{-6}$ to compare with the fitness upper bound in SDENUA. Fig. 8 shows the priori spectra and the unfolding results of the four spectra in the IAEA 403 report [18].

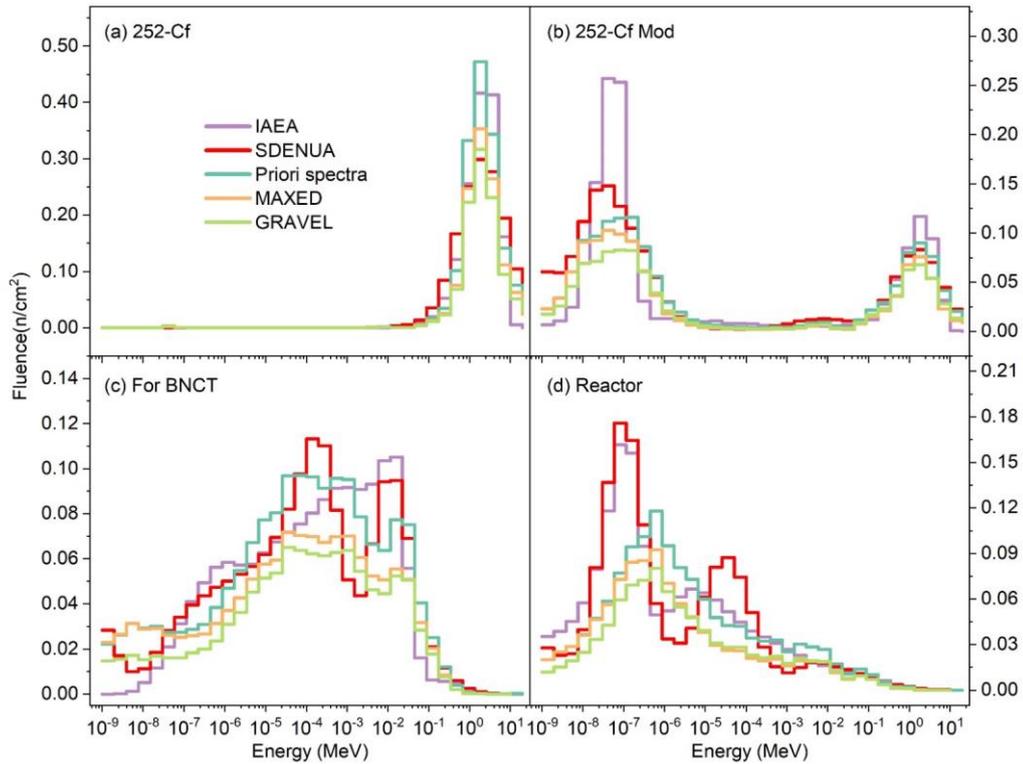

**Fig. 8** Neutron spectra unfolding results (with average over 20 independent runs in SDENUA): (a) spectrum of the isotope source of Cf-252, (b) spectrum of Cf-252 source after being moderated, (c) spectrum used for BNCT, and (d) spectrum of a certain reactor from Germany. The uncertainties for the calculated spectra are less than 5%, which were calculated as follows in the Appendix.

Except for Fig. 8a, the unfolding results of the other three energy spectra show obvious errors comparing with the reference spectra in the energy range from $10^{-9}$ MeV to $10^{-7}$ MeV. As shown in Fig. 2, the response functions of energies from $10^{-9}$ MeV to $10^{-7}$ MeV have severely overlapped, which weakens the unfolding power of the response functions in this energy range. Due to the integral fitness function, as shown in Figs. 8c and 8d, although the final optimal solutions fluctuate in a large range around the reference spectra, the fitness of the final solutions still reached $10^6$. However, this also means that the solution space that can meet the error constraint is expanded, and fewer iterations for searching are required, which can be supported in Fig. 6.

The results given by UMG 3.1 are interesting. Although an excellent priori default energy spectrum is provided, the agreement between the results and the reference spectra is poor. This may be due to neutron counts and response matrix input with uncertainties, and the UMG 3.1 unfolded the neutron spectra following the inputs strictly.

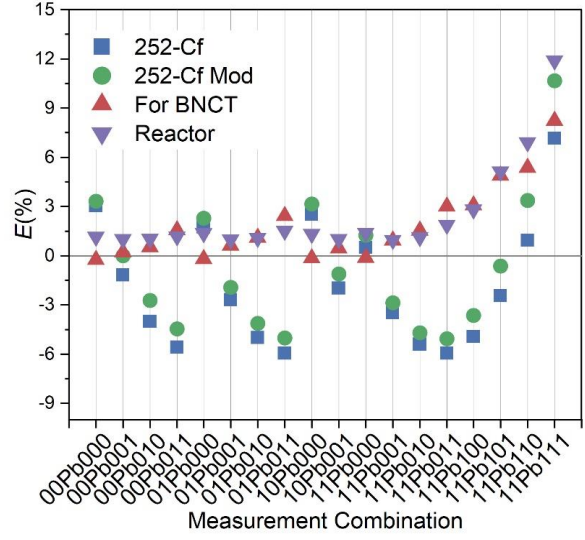

**Fig. 9** The uncertainties of the unfolded results. The uncertainties for the calculated neutron counts are less than 5%, which were calculated as follows in the Appendix.

The uncertainties of the unfolded results are from both the uncertainty terms of the measured neutron counts and the response matrix. Fig. 9 shows that the final results are given by SDENUA with errors from -6% to 12%, which calculated follow

$$E_j = \frac{C_j^{cal} - C_j^{meas}}{C_j^{meas}} \times 100, (16)$$

where $E_j$ is the error of the $j$th measurement combination, $C_j^{cal}$ is the calculated count of $j$th measurement combination, and $C_j^{meas}$ is the same as Eq. (1).

Although the fitness value of the final solution is limited to $10^6$, there is still a 12% error between the unfolding result and the exact measured count due to the introduction of the uncertainties both of the measured counts and the response matrix. In the neutron unfolding problem, we usually pay more attention to obtaining a suitable spectrum instead of a spectrum with ultra-small counting errors [15]. Because the ultra-small errors do not produce an ideal analytical solution of the first kind Fredholm integral problem biased on the integral fitness function [9]. So, combining Fig. 8 and Fig. 9, we hold the view that the errors of the unfolded results are an acceptable level of accuracy.

The fitness function used in this work is by matching the integral quantities of input neutron counts and calculated neutron counts. This will trigger a discussion about the mechanism in the selection operation. On the positive side, the candidate individual that has failed to evolve will be denied entry into the population in time to avoid the reduction of the population quality. On the negative side, first, as shown in Fig. 3, the fitness value of an individual (solution) is higher, and the corresponding QS (the error indicator) may be low, so that the error between the solution and the actual energy spectrum may be worse even with a high fitness value. Besides, an individual is composed of multiple genes, a few excellent genes will be immediately eliminated from the population due to the low fitness of the candidate individuals. As a result, even a potential candidate individual will be immediately affected because of the low fitness value.

As far as the SDENUA established in this work is concerned, the QS values of the optimal solution of "Cf-252" and " Cf-252 Mod" in Fig. 7 are worse than that of "For BNCT" and "Reactor" when the final optimal solution reaches $10^6$. In other words, it shows that limiting the fitness of the final optimal solution may sacrifice the unfolding accuracy of the "Cf-252" and " Cf-252 Mod" energy spectra. We hypothesize that one of the reasons for this phenomenon is that the "Cf-252" and " Cf-252 Mod" energy spectra contain fewer energy groups with non-zero fluences, which is equivalent to reducing the number of effective constraint items for calculated neutron counts in the convolution process. In summary, for such an energy spectrum containing a large number of 0 fluence energy groups, $10^6$ as the fitness upper of the optimal solution may not large enough, but considering other types of the energy spectrum, we can only make a trade-off.

Regarding the priori spectrum in the neutron energy spectrum unfolding problem, different authors hold different views. On the one hand, the developers of UMG 3.1 mentioned in the literature [12] that the problem of neutron energy spectrum unfolding should be based on an excellent priori default energy spectrum. They believed that adding the physical information of the neutron energy spectrum helps to obtain a more accurate unfolding result. On the other hand, the authors who use artificial intelligence algorithms to believe that it is difficult to estimate an excellent priori spectrum in some cases. So, the dependence on the priori energy spectrum should be reduced. The former is explained from the perspective of physics while the latter from mathematics. From our point of view, we believe that it is hard to obtain an accurate solution only from the perspective of solving the first kind Fredholm integral problem, although the literature [14-16] has been done a lot of meaningful works. As far as the algorithm based on fitness function is concerned, some physical information of the neutron energy spectrum can be added to the fitness function as a constraint instead of asking for an excellent priori spectrum, such as the continuity of the neutron spectrum. This measure may have a positive effect on the unfolding, which requires a lot of experiments and in-depth research, and these works will be carried out in the future of our work.

## 4 Conclusion

Finally, conclusions and further work are summarized. The self-adaptive differential evolution neutron spectrum unfolding algorithm (SDENUA) has achieved promising results for the Water-pumping-injection Multi-layered concentric sphere Neutron Spectrometer (WMNS). In the mutation operation, the information of High-fitness individuals and failure individuals was used to improve the guidance of the evolution direction of the population, so that the fitness of the final optimal solution can quickly reach $10^6$ to output an acceptable solution, thereby improving the quality of the solution and shortens the running time. Historical experience information was adopted to perform the self-adaptive control of scaling factor and crossover probability. The constructed self-adaptive difference algorithm was used to unfold the readings simulated from the (1) spectrum of Cf-252 and (2) its spectrum after being moderated, (3) a spectrum used for BNCT, and (4) a spectrum from a reactor in the IAEA 403 report. The unfolding spectra have good agreement with the reference spectra than that of UMG 3.1, which demonstrate that in the absence of a priori default spectrum, and the uncertainties of the measured neutron counts and the response matrix, the unfolded results are an acceptable level with errors under 12%.

# 5 Further work

How to introduce the physical information of the energy spectrum to improve the results is also interesting. Besides, in order to better verify the performance of the algorithm in actual applications, more experiments will be carried out. These tasks will be implemented in the future.

**Acknowledgments**

This work was supported by the National Natural Science Foundation of China (No. 41774120) and by the Sichuan Science and Technology Program (No. 2018TJPT0008, No. 2019YFG0430).

**Appendix**

In this work, the type A standard uncertainty [27] are used and calculated follow

$$\Delta U^A = \sqrt{\frac{\sum_{1}^{n}(x_i - \bar{x})^2}{n(n-1)}},$$

where $\Delta U^A$ is uncertainty of type A with the same unit as the variable $x$, $x_i$ is the calculated value of variable $x$, in the spectra uncertainty estimation ($x_i$ is the calculated neutron fluence of the $i$th runs, in the neutron counts uncertainty estimation, $x_i$ is the calculated neutron count of the $i$th runs), $\bar{x}$ is the mean of the variable $x$, $n$ is the number of independent runs. And the uncertainty of type A with percent is estimated follow

$$\Delta U_p^A = \frac{\Delta U^A}{x^{exac}} \times 100,$$

where $x^{exac}$ is the exact value of the variable $x$.